\documentclass[12pt,a4paper]{article}
\usepackage{amsmath,amssymb}
\usepackage{graphicx}
\usepackage{color}
\usepackage{cite}
\newcommand{\eqn}[1]{&\hspace{-0.6em}#1\hspace{-0.6em}&}

\begin{document}
\setlength{\baselineskip}{18pt}
\begin{titlepage}

\vspace*{1.2cm}
\begin{center}
\hfill TU-1019, IPMU16-0038\\
\vskip .75in

{\Large\bf 

Renormalization Group Improved Higgs Inflation \\
with a Running Kinetic Term
}
\end{center}
\lineskip .75em
\vskip 1.5cm

\begin{center}
{\large Fuminobu Takahashi}$^{1,2}$ and
{\large Ryo Takahashi}$^3$ \\

\vspace{1cm}

$^1${\it Department of Physics, Tohoku University, Sendai 980-8578, Japan}\\
$^2${\it Kavli Institute for the Physics and Mathematics of the Universe (WPI), TODIAS,\\ University of Tokyo, Kashiwa 277-8583, Japan}\\
$^3${\it Graduate School of Science, Tohoku University, Sendai 980-8578, Japan}\\

\vspace{10mm}
{\bf Abstract}\\[5mm]
{\parbox{13cm}{\hspace{5mm}
We study a Higgs inflation model with a running kinetic term, taking account of
the renormalization group evolution of relevant coupling constants.
Specifically we study two types of the running kinetic Higgs inflation, where 
the inflaton potential is given by the quadratic or linear term potential in a 
frame where the Higgs field is canonically normalized. We solve  the 
renormalization group equations at two-loop level and calculate the scalar 
spectral index and the tensor-to-scalar ratio. We find that, even if the 
renormalization group effects are included, the quadratic inflation is ruled out
 by the CMB observations, while the linear one is still allowed.
}}
\end{center}
\end{titlepage}

\section{Introduction}
A Higgs boson was discovered at the LHC~\cite{Chatrchyan:2013lba,CMS}, which 
completed the last missing piece of the standard model (SM). While the 
properties of the Higgs boson are consistent with the expected ones for the SM
Higgs boson within experimental uncertainties, a deeper understanding of the 
Higgs sector might be a key to unravel various phenomena which cannot be 
explained within the SM,  such as dark matter, baryogenesis, etc. In fact, the 
Higgs field can play a role of the inflaton field responsible for cosmic 
acceleration in the early Universe~\cite{Guth:1980zm,Starobinsky:1980te,Sato:1980yn,Linde:1981mu,Albrecht:1982wi}.
There have been proposed a variety of Higgs inflation models (see e.g. 
Refs.~\cite{CervantesCota:1995tz,Bezrukov:2007ep,Bezrukov:2009db,Einhorn:2009bh,Germani:2010gm,Ferrara:2010yw,Lee:2010hj,Ferrara:2010in,Nakayama:2010sk,Kamada:2010qe,Masina:2011aa,Masina:2011un,Hertzberg:2011rc,Kamada:2012se,Allison:2013uaa,Hamada:2013mya,Nakayama:2014koa,Cook:2014dga,Hamada:2014iga,Bezrukov:2014bra,Haba:2014zda,Haba:2014zja,Hamada:2014xka,He:2014ora,Hamada:2014raa,George:2015nza,DiVita:2015bha,Ge:2016xcq} and references therein). 

The successful Higgs inflation requires a rather flat potential. At large field 
values, the SM Higgs potential at tree-level is approximately given by a quartic
 potential with a coefficient of order $0.1$, which, however, is too steep to 
drive successful inflation and the density perturbations would be too large if 
we extrapolate the potential up to super-Planckian values. Therefore, the Higgs 
potential must be somehow modified at large field values for successful 
inflation. In this paper, we consider the Higgs inflation with a running kinetic
 (RK) term~\cite{Nakayama:2010sk,Nakayama:2014koa}, where the kinetic term of 
the Higgs field is allowed to depend on the Higgs field itself. Such inflation 
models with a RK term as well as their phenomenological and cosmological 
implications were studied in detail in 
Refs.~\cite{Takahashi:2010ky,Nakayama:2010kt}.

In the simplest realization, the kinetic term of the Higgs field is given 
by~\cite{Nakayama:2010sk,Nakayama:2014koa}
\begin{eqnarray}
{\cal L}_K = \frac{1}{2} \left(1 + \xi h^2\right ) (\partial h)^2,
\label{simple}
\end{eqnarray}
where $\xi$ is a numerical coefficient and $h$ is the Higgs field. Thus, at 
sufficiently large field values, the quartic potential becomes a simple 
quadratic potential when expressed in terms of a canonically normalized field, 
$\hat{h} \propto h^2$. In contrast to the original Higgs inflation with a 
non-minimal coupling to gravity~\cite{Bezrukov:2007ep}, the predicted 
tensor-to-scalar ratio, $r$, is relatively large, which is therefore tightly 
constrained by the recent Planck and BICEP2/Keck Array 
experiments~\cite{Ade:2015lrj,Array:2015xqh}.

It is known that the Higgs potential is modified by  the running effects of 
coupling constants, and it is especially sensitive to the top Yukawa coupling. 
So far, the Higgs inflation model with a RK term was studied only at tree-level,
 and therefore, it was not clear if the simplest realization of the model is 
still allowed by observations, once one properly takes account of the running 
effects. 

In this paper we take into account the running effects of relevant coupling 
constants under renormalization group equations (RGEs) at two-loop level with 
the experimentally observed value of the Higgs mass $M_h=125.6$ GeV, and study 
the inflation dynamics of the Higgs inflation with a RK term. Then we calculate 
the scalar spectral index $n_s$ and the tensor-to-scalar ratio, and compare them
with the observations. We will show that the recent Planck and BICEP2/Keck Array
 results rule out the simplest realization where the inflaton potential is given
 by the quadratic term, even if one takes account running effects of coupling 
constants under RGEs. We also study another realization of the RK Higgs 
inflation where the inflaton potential is given by a linear term, which is shown
 to be consistent with the observations even if the running effects are taken 
into account.

The rest of this paper is organized as follows: In the next section, we 
investigate the RK Higgs inflation with a quadratic or linear term, taking 
account of the RGEs for the relevant coupling constants. The last section is 
devoted to discussion and conclusions. In the Appendix we give $\beta$-functions
 up to two-loop level for the RGEs of relevant coupling constant.

\section{RGE improved RK Higgs inflation}

In this section we investigate two types of the RK Higgs inflation model, 
including the running effects of relevant coupling constants by solving their 
RGEs. We first study the simplest realization where the inflaton potential is 
given by the quadratic term, and compare the predicted $n_s$ and $r$ with 
observations. We also study another realization where the inflaton potential is 
a linear term.

\subsection{Quadratic model}

The base model of the RK inflation is given 
by~\cite{Takahashi:2010ky,Nakayama:2010kt,Nakayama:2010sk,Nakayama:2014koa,Hertzberg:2011rc}
 \begin{eqnarray}
  \mathcal{L}=\frac{1}{2}(1+\xi\phi^2)(\partial\phi)^2-V(\phi),
  \label{LQ}
 \end{eqnarray}
where $\xi$ is a positive numerical coefficient much larger than unity, $\phi$ 
is the inflaton, and $V(\phi)$ is the inflaton potential. Here and in what 
follows we adopt the Planck units where the reduced Planck mass $M_{\rm pl}$ is set to be 
unity. The canonically normalized inflaton field is given by 
$\hat{\phi}\sim\sqrt{\xi}\phi^2$ at $\phi\gtrsim1/\sqrt{\xi}$, because the 
kinetic term is dominated by the $\xi \phi^2$ term at large field values. Thus, 
the scalar potential changes its form above the critical value,
 \begin{eqnarray}
V(\phi) \rightarrow V(\hat{\phi}{}^{1/2}/\xi^{1/4}).
  \label{Vc}
 \end{eqnarray}
For instance, if the scalar potential in the original frame contains the quartic
 term $\phi^4$, it turns into the  quadratic one, $\hat{\phi}^2/\xi$, at 
$\phi\gtrsim1/\sqrt{\xi}$ when expressed in terms of  the canonically normalized
 field. Thus, the inflaton potential becomes flatter at large field values due 
to the RK term.

In order to have successful large-field inflation, one has to keep a handle on
the interactions of the inflaton, especially the large
coupling in the kinetic term, at super-Planckian field values. 
One possible way is to impose a shift symmetry
on $\phi^2$,
\begin{equation}
\phi^2 \to \phi^2 + C,
\end{equation}
where $C$ is a (real) transformation parameter. 
Then, the canonically normalized inflaton field ${\hat \phi}$ is
necessarily proportional to $\phi^2$ at sufficiently large field values, as the form
of the kinetic term is dictated by the symmetry. In this case, the largeness of $\xi$ 
can be understood because the ordinary kinetic term as well as the potential
breaks the shift symmetry, and they should be accompanied by a small order parameter.
Once one normalizes the inflaton field $\phi$ so that it is canonically normalized at $\phi = 0$,
the small order parameter is translated to the large $\xi$.

We apply  the above base model to the SM Higgs field $H$, 
and the relevant part of the Lagrangian is given by
 \begin{eqnarray}
  \mathcal{L}\supset\frac{1}{2}(1+\xi  h^2)(\partial h)^2
                    -\frac{\lambda}{4}( h^2-v^2)^2,
 \label{L2}
\end{eqnarray}
where $h$ is the physical Higgs field. The largeness of $\xi$ in 
Eq.~(\ref{LQ}) can be explained by the smallness  breaking
of the shift symmetry,
 \begin{eqnarray}
  |H|^2\rightarrow |H|^2+C,
  \label{shift}
 \end{eqnarray}
which keeps the potential under control at large field values. For 
$h\gg 1/\sqrt{\xi}$, the relevant terms in Eq.~(\ref{L2}) 
are given by 
 \begin{eqnarray}
 \label{m2phi2}
  \mathcal{L}\approx\frac{1}{2}(\partial\hat{h})^2- \frac{\lambda}{\xi} \hat{h}^2,
 \end{eqnarray}
with the canonically normalized field $\hat{h}\equiv \sqrt{\xi} h^2/2$. Thus, 
the quadratic chaotic inflation takes place if $\hat{h}$ is initially located at
 large field values. 

The largeness of $\xi$ in Eq.~(\ref{LQ}) may raise doubts in the validity of the
 RK inflation. Namely, the high energy scattering amplitudes for $\phi \phi \to 
\phi \phi$ imply that the system enters into a strongly-coupled regime at energy
 scales much below the Planck scale. On the other hand, as we have seen above, 
the large $\xi$  arises from a small shift-symmetry breaking parameter, and 
clearly the inflation dynamics is described in a weakly-coupled regime. The 
apparent  tension can be understood by noting that the perturbative unitarity 
bound is actually field-dependent~\cite{Bezrukov:2011sz}, and that the kinetic 
term grows as the inflaton field increases in a controlled way thanks to the 
shift symmetry. First, the cutoff scale due to the non-minimal kinetic term is
\begin{equation}
\Lambda(h) \sim \left\{
\begin{array}{cc}
\displaystyle{M_{\rm pl}/\sqrt{\xi}} & {\rm for~} h \lesssim \frac{M_{\rm pl}}{\sqrt{\xi}} \\
\displaystyle{\sqrt{\xi} h^2/M_{\rm pl}}& {\rm for~} h \gtrsim \frac{M_{\rm pl}}{\sqrt{\xi}} \\
\end{array}
\right.,
\end{equation}
while the cutoff due to purely gravitational interactions is of order the Planck mass, 
and we explicitly show the dependence of the Planck mass for clarity. 
Therefore, during inflation when  $h$ is larger than $M_{\rm pl}/\sqrt{\xi}$,
the cutoff is of order the Planck mass, and there is no problem in using the effective field theory (\ref{L2})
to describe the inflaton dynamics. The situation is quite analogous to the Higgs inflation with a non-minimal coupling
to gravity. Indeed, as noted in Ref.~\cite{Bezrukov:2014bra}, for a certain range of the Higgs field, 
the quadratic potential can also be obtained if one has the following Higgs-gravity coupling,
\begin{equation}
S = \int d^4 x \sqrt{-g} 
\left(\frac{M_{\rm pl}^2}{2}  f(h) R
+ \frac{1}{2} g^{\mu \nu} \partial_\mu h \partial_\nu h - V(h) \right)
\end{equation}
with
\begin{equation}
f(h) \;=\; 
1 + \frac{h^2}{\sqrt{6} \epsilon M_{\rm pl}^2}.
\end{equation}
In the Einstein frame, the kinetic term of $h$ is
\begin{equation}
{\cal L}_K \;=\; \frac{1}{2} \left(\frac{1}{f}+\frac{3 f'{}^2 M_{\rm pl}^2}{2f^2} \right)  \partial_\mu h \partial^\mu h,
\end{equation}
where the prime denotes the derivative with respect to $h$. For $\epsilon M_{\rm pl} \lesssim h \lesssim \sqrt{\epsilon} M_{\rm pl}$,
the above kinetic term can be approximated by
\begin{equation}
{\cal L}_K \;\simeq\; \frac{1}{2} \left(1+ \epsilon^{-2} \frac{h^2}{M_{\rm pl}^2}   \right)  \partial_\mu h \partial^\mu h
\simeq \frac{1}{2}  \partial_\mu {\hat h} \partial^\mu {\hat h}
\end{equation}
which is the same as Eq.~(\ref{L2}) if $\xi = 1/\epsilon^2$. This implies that 
the RK Higgs inflation is equivalent to the Higgs inflation with the above 
Higgs-gravity coupling for ${\hat h} \lesssim M_{\rm pl}$, and that the RGE 
effects of the relevant couplings can be similarly taken into account. At 
super-Planckian field values, the inflaton potential is still given by the 
quadratic potential in our model because of the shift symmetry. Secondly, the 
form of the kinetic term is determined by the shift symmetry, and therefore, it 
is robust against radiative corrections. In particular,  the canonically 
normalized inflaton $\hat{h}$ is always proportional to $h^2$ at sufficiently 
large field values, and this relation is not spoiled by including radiative 
corrections. This implies that the inflaton potential can be well approximated  
by a quadratic mass term for a canonically normalized inflaton, as long as the 
scalar potential is dominated by a quartic term in Eq.~(\ref{L2}). When one 
takes account of the running of coupling constants, there is a subtlety in the 
renormalization prescription, but this does not affect the validity of the RK 
inflation. In particular, our main result remains unchanged, and the reason
for this will become clear shortly.

Now we study the above inflation model, including the running effects by solving
 the RGEs at two-loop level for  relevant coupling constants, which enables 
precise comparison between predictions and observations. The corresponding RGEs 
are given by
\begin{eqnarray}
(4\pi)^2\frac{dX}{dt}=\beta_X,
\end{eqnarray}
where $X$ collectively denotes the SM gauge coupling constants $g_i~(i=1,2,3)$, 
the top Yukawa coupling $y_t$, and the Higgs quartic coupling $\lambda$. $t$ is 
defined by $t\equiv\ln(\mu/1~{\rm GeV})$ where $\mu$ is the renormalization 
scale. We numerically solve the RGEs within the renormalization scale of 
$M_Z\leq\mu\leq m_{\rm pl}$ where $M_Z$ is the $Z$ boson mass $M_Z=91.2$ GeV and 
$m_{\rm pl}$ is the Planck mass $m_{\rm pl}=1.22\times10^{19}$ GeV. The 
$\beta$-functions for the coupling constants are given in the Appendix. In the 
analysis, we take the Higgs mass to be $M_h=125.6$ GeV.

The metric perturbations are usually characterized by the scalar spectral index 
$n_s$ and the tensor-to-scalar ratio $r$. In our analysis we adopt the 
first-order expressions, $n_s=1-6\varepsilon+2\eta$ and  $r=16\varepsilon$, 
respectively. Here the slow roll parameters (in the Planck units) are defined as
\begin{eqnarray}
\varepsilon=\frac{1}{2}\left(\frac{V'}{V}\right)^2,~~~
\eta=\frac{V''}{V},
\end{eqnarray}
where $V'\equiv dV/d\hat{\phi}$ and $V''\equiv d^2V/d\hat{\phi}^2$. Here 
$\hat{\phi}$ is the canonically normalized inflaton (Higgs) field. The inflation
 ends when either of $\varepsilon$ or $|\eta|$ exceeds the unity. The e-folding 
number $N$  is given by 
\begin{eqnarray}
N=\int_{\hat{\phi}_{\rm end}}^{\hat{\phi}_0}\frac{V}{V'}d\hat{\phi},
\end{eqnarray}
where $\hat{\phi}_0$ and $\hat{\phi}_{\rm end}$ are the initial and final field 
values during the inflation, respectively. We have evaluated $n_s$ and $r$ for 
the e-folding number $N$ between $50$ and $60$, and obtained the following 
results,
\begin{eqnarray}
0.960\lesssim n_s\lesssim 0.967,~~~
0.132\lesssim r\lesssim 0.159~~~\mbox{for}~~~
50\leq N\leq60,
\end{eqnarray}
where we take the scalar amplitude as $A_s=(2.196_{-0.060}^{+0.051})\times10^{-9}$, 
and set the prior on the parameter $\xi$ as $\xi \gtrsim 1.36\times10^5$ and the
 top mass as $M_t\leq 171.2$ GeV, for numerical stability. We have found that 
the values of $n_s$ and $r$ are rather robust against the renormalization-group 
effects, and their variations  with respect to the 
top mass are very small, and approximately given by 
$\delta n_s\simeq2\times10^{-5}$ and $\delta r\simeq10^{-4}$ for a fixed e-folding
 number. Here we vary the top quark mass as 
$168\,{\rm GeV}<M_t<171.2\,{\rm GeV}.$ Thus, even if the running  effects of 
coupling constants under RGEs are taken into account, the base RK Higgs 
inflation model with a quadratic potential is ruled out by the recent 
Planck~\cite{Ade:2015lrj} and BICEP2/Keck Array~\cite{Array:2015xqh} results, 
which placed an upper bound on $r$ as $r < 0.07$ at 95$\%$ CL.

The reason for the robustness of $n_s$ and $r$ can be understood as follows. 
Even though the Higgs quartic coupling is sensitive to the top quark mass, as 
long as it is positive during inflation, the inflaton potential is still well 
approximated by the quadratic potential (plus small logarithmic corrections) 
except for the top quark mass $M_t \gtrsim 170$\,GeV. The values of $n_s$ and 
$r$ are sensitive only to the shape of the potential, and they do not depend on 
the overall normalization of the potential, which is determined by the size of 
the quartic coupling at large field values, and that is why their values are 
robust against including the RGE effects. On the other hand, the value of $\xi$ 
is determined by the normalization of the curvature perturbations, and 
therefore, it is sensitive to the absolute value of $\lambda$ and $M_t$. 
 
\begin{figure}[t]
\begin{center}
\includegraphics[scale=1,bb=0 0 265 265]{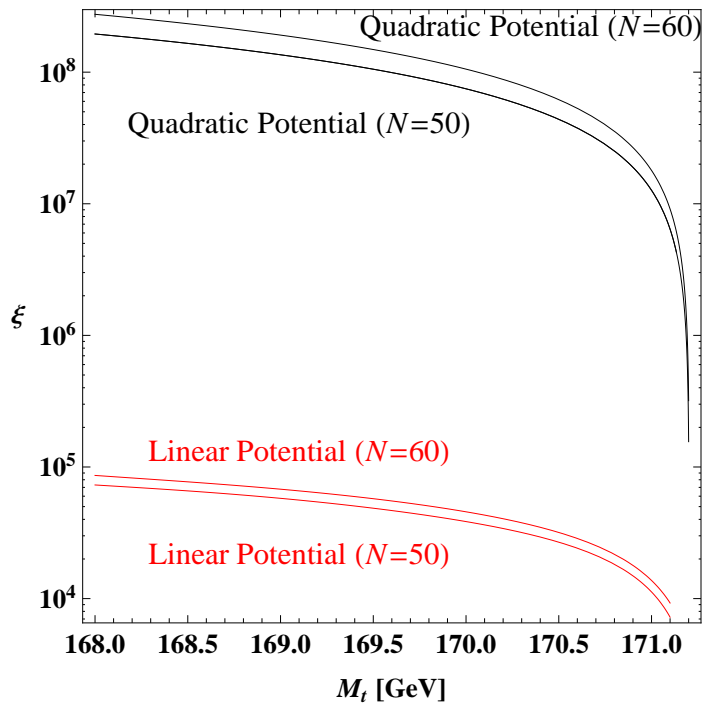}
\end{center}
\caption{The dependence of $\xi$ on the top quark mass $M_t$ in the RGE improved
 Higgs inflation with RK term models. Black (top) and red (bottom) lines 
correspond to models with quadratic and linear potentials, respectively.}
\label{fig1}
\end{figure}

In Figure \ref{fig1} we show the dependence of $\xi$ on $M_t$: the 
black line indicates the dependence in this quadratic potential model. As the 
top quark mass increases up to $M_t \simeq 170$\,GeV, the value of $\lambda$ 
becomes smaller and smaller, and in order to generate the curvature 
perturbations of the right magnitude, the value of $\lambda$ and $\xi$ must 
satisfy
\begin{eqnarray}
\frac{\lambda}{\xi} \simeq  2 \times 10^{-11},
\end{eqnarray}
at a scale relevant for inflation. This explains the behavior of $\xi$ in 
Figure~\ref{fig1}. 

For $M_t \gtrsim 170$\,GeV, the quartic coupling $\lambda$ becomes even smaller,
 and at a certain point, the inflaton potential develops local maximum and 
minimum. In order for the inflaton to roll down to the electroweak vacuum, its 
initial position $h_{\rm ini}$ must be smaller than $h_{\rm max}$ at which the 
inflaton potential takes the local maximum. Thus, the quartic coupling $\lambda$
 during inflation cannot be arbitrarily small, and it is bounded 
below~\cite{Hamada:2014iga},
\begin{equation}
\lambda(h) \gtrsim \lambda(h_{\rm max}) \simeq 5 \times 10^{-6}.
\end{equation}
Thus, in order to explain the observed density perturbations, $\xi$ is also 
bounded below as $\xi \gtrsim 3 \times 10^{5}$, which is consistent with 
Figure~\ref{fig1}. As $M_t$ further increases, the location of the local maximum
 becomes smaller than $M_{\rm pl}$. As a result, one of the slow-roll parameters, 
$|\eta|$, exceeds unity, and the slow-roll inflation becomes difficult to 
realize. This explains why the allowed region rapidly shrinks at $M_t \gtrsim 
170$\,GeV.
 
\subsection{Linear  model}
Let us now consider a RK Higgs inflation with a linear potential. The Lagrangian
is given by
 \begin{eqnarray}
  \mathcal{L}=\frac{1}{2}(1+\xi^3\phi^6)(\partial\phi)^2-\frac{\lambda}{4}\phi^4.
  \label{LL}
 \end{eqnarray}
At large field values, $\phi\gtrsim1/\sqrt{\xi}$, the canonically normalized 
inflaton field is $\hat{\phi}\sim\xi^{3/2}\phi^4$, and the effective inflaton 
potential becomes a linear potential. Thus, the inflaton potential becomes 
flatter compared to the quadratic one, and the tensor-to-scalar ratio is 
expected to be suppressed.

In order to apply  the above model to the SM Higgs field $H$, we introduce a 
shift symmetry
 \begin{eqnarray}
  |H|^4\rightarrow |H|^4+C,
 \end{eqnarray}
and introduce the ordinary kinetic term  and  the Higgs potential as small 
explicit breaking of the symmetry. 
We consider the following Lagrangian,
 \begin{eqnarray}
  \mathcal{L}\supset\frac{1}{2}(1+\xi^3 h^6)(\partial h)^2
                    -\frac{\lambda}{4}( h^2-v^2)^2,
 \end{eqnarray}
where $h$ is the physical Higgs. 
At large field values as $h\gg \xi^{-3/4}$, the relevant terms in Eq.~(\ref{L2}) 
are given by 
 \begin{eqnarray}
  \mathcal{L} \approx \frac{1}{2}(\partial\hat{h})^2 - \frac{\lambda}{\xi^{3/2}} \hat{h},
 \end{eqnarray}
with the canonically normalized field $\hat{h}\equiv \xi^{3/2} h^4/4$. Thus, the
 linear inflation is realized.

We solve the inflaton dynamics, taking into account of the RGE evolution of the 
relevant couplings. Then we obtain the following results for $n_s$  and $r$, 
model leads
\begin{eqnarray}
0.970\lesssim n_s\lesssim 0.975,~~~
0.066\lesssim r\lesssim 0.079~~~\mbox{for}~~~
50\leq N\leq60,
\end{eqnarray}
where we take the scalar amplitude as $A_s=(2.196_{-0.060}^{+0.051})\times10^{-9}$, 
and set a prior on the parameter $\xi$ as  $\xi \gtrsim 7.01\times10^3$, and the
 top mass as $M_t\leq 171.1$ GeV for numerical stability. We have found that, 
similar to the quadratic inflation, $n_s$ and $r$ are robust against including 
the RGE evolution of the relevant couplings, since they are sensitive to the
inflaton potential shape only, which changes only logarithmically due to the RGE
 effects. Specifically we have found the values of $n_s$ and $r$ vary with 
respect to the top quark mass as $\delta 
n_s\simeq2\times10^{-5}$ and $\delta r\simeq10^{-4}$ for a fixed e-folding number.
 Here we vary the top quark mass as $168\,{\rm GeV} < M_t < 171.1\,{\rm GeV}.$ 
Thus, this model is consistent with the CMB 
observation~\cite{Ade:2015lrj,Array:2015xqh}. The value of $\xi$, on the other 
hand, depends on the top quark mass through the RGE running of the Higgs quartic
 coupling.  We show the dependence as red (bottom) lines  in Fig.~\ref{fig1}. As
 expected, the value of $\xi$ decreases as the Higgs quartic asymptotes to zero 
at $M_t \simeq 171$\,GeV.

\section{Discussion and Conclusions}
In our analysis we set an upper bound of the top quark mass $M_t\lesssim 171$ GeV.
 This is because the Higgs quartic coupling must be positive at a scale relevant
 for inflation. The upper bound on the top quark mass, taken at a face value, is
 out of the experimentally measured value ($M_t=173.34\pm0.76$ 
GeV~\cite{ATLAS:2014wva}). This discrepancy, however, is not so 
problematic~\cite{Hamada:2014xka,Haba:2014qca}, because the experimental value 
is  derived as an invariant mass for the final states of color singlets to fit 
data while $M_t$ in the above calculation of the RGEs is the pole mass. There 
might be a few GeV discrepancy between the singlet final state and the color 
octet $t\bar{t}$ pair, which is dominant at the hadron 
collider~\cite{Horiguchi:2013wra}. That said, it is also possible to have 
successful RK Higgs inflation for the top quark mass close to the experimental 
central value, by introducing other particles such as a singlet scalar dark 
matter  with a TeV mass and heavy right-handed neutrinos. Since the new 
particles modify the $\beta$-functions of the coupling constants, the Higgs 
quartic coupling remains positive at a higher scale for a heavier top 
mass~\cite{Haba:2014zda,Hamada:2014xka,Haba:2014zja,Haba:2013lga,Haba:2014sia}. 
In such an extension, the model can also explain dark matter  by the singlet 
scalar, and small active neutrino masses and baryon asymmetry of the Universe by
 heavy right-handed neutrinos, respectively. The extension for this model will 
be given in another publication.

The RK Higgs inflation with a quadratic potential predicted a too large 
tensor-to-scalar ratio. In this paper we have considered another form of the 
kinetic term so that the effective potential is a linear term. Another way to 
modify the inflaton potential is to add higher dimensional operators. Then the 
inflaton potential is given by a polynomial instead of a monomial 
one~\cite{Nakayama:2013jka}.

Our Universe has experienced  an accelerating expansion in the early epoch, i.e.
 the so-called inflation. Yet it remains unknown what the inflaton is. The 
recently discovered Higgs particle is the unique elementary scalar particle 
among the known particles, and it might play a role of the inflation. The idea 
of the Higgs inflation has attracted much attention, and studied extensively in 
the literatures. In this paper, we have focused on the so-called RK Higgs 
inflation, where the kinetic term is allowed to depend on the Higgs field itself
 and therefore the effective potential becomes flatter at large field values. So
 far, the RK Higgs inflation was studied only at tree level, and therefore, it 
was not clear if its simplest realization (\ref{simple}) is still allowed by the
 CMB observations.  

In this paper we have first considered the running effects of various coupling 
constants under the RGEs on two types of the RK Higgs inflation. One of the 
models has the quadratic potential and the other has a linear potential. We have
 shown that the quadratic potential model leads $0.960\lesssim n_s\lesssim 
0.967$ and $0.132\lesssim r\lesssim 0.159$ for the e-folding number $N$ between 
$50$ and $60$, and that the values of $n_s$ and $r$ are quite robust against 
including the RGE effects. In fact, their variations are $\delta 
n_s\simeq2\times10^{-5}$ and $\delta r\simeq10^{-4}$ for a fixed e-folding number 
and the top quark mass $168\,{\rm GeV} < M_t < 171.2\,{\rm GeV}.$ Thus, the RK 
Higgs inflation model with a quadratic potential is ruled out by the recent 
Planck and BICEP2/Keck Array ($r<0.07$)~\cite{Ade:2015lrj,Array:2015xqh} result 
even if one takes account running effects of coupling constants under RGEs. 
Next, we have discussed the model with a linear potential. Similarly, we have 
shown that $n_s$ and $r$ are robust against including the running effects, and 
they are given by $0.970\lesssim n_s\lesssim 0.975$ and $0.066\lesssim 
r\lesssim0.079$ with $\delta n_s\simeq2\times10^{-5}$ and $\delta r\simeq10^{-4}$,
 which are consistent with the current observations. While $n_s$ and $r$ are not
 significantly modified by the RGE effects, the coefficient $\xi$ in the kinetic
 term is sensitive to the running effects. This is because, while $n_s$ and $r$ 
are sensitive only to the inflaton potential shape, the magnitude of the 
curvature perturbation is also sensitive to the overall scale of the inflaton 
potential. As a result, $\xi$ varies with respect to the top quark mass, and we 
have shown that it decreases toward the critical value of the top quark mass at 
which the Higgs quartic coupling becomes extremely small at a scale relevant for
 inflation.

One of the virtues of the Higgs inflation with a RK term is that the predicted 
value of tensor-to-scalar ratio is larger than the Higgs inflation with a 
non-minimal coupling to gravity; the typical size of $r$ is of ${\cal 
O}(0.01-0.1)$. Future observations of the CMB B-mode polarization will refute or
 support the RK Higgs inflation with a linear term or even flatter or polynomial
 potential.

\section*{Acknowledgments}
This work was supported by the Grant-in-Aid for Scientific Research on 
Scientific Research A (No.26247042 [FT]), Scientific Research B (No. 26287039 
[FT]), Young Scientists B (No. 24740135 [FT]) and Innovative Areas (No. 23104008
 and No.15H05889 [FT]). This work was supported by World Premier International 
Research Center Initiative (WPI Initiative), MEXT, Japan. 

\appendix
\section*{Appendix: {\boldmath $\beta$}-functions}

We give the $\beta$-functions of the relevant coupling constants at two-loop 
level.\footnote{Another prescription to obtain the RGEs at one-loop level has 
been presented in~\cite{George:2015nza}. In the work, the authors demand that 
the Higgs inflation model can be expanded by small parameters in large and mid 
field regime, and all loop corrections can be absorbed in counterterms at each 
loop level.} The $\beta$-functions for the gauge coupling constants are
\begin{eqnarray}
\beta_{g_1} \eqn{=} 
\frac{3}{5}\left(\frac{81+s}{12}\right)g_1^3 
+\frac{1}{16 \pi^2}\frac{3}{5}g_1^3 
\left[\frac{199}{30}g_1{}^2+\frac{9}{2}g_2^2+\frac{44}{3}g_3^2-\frac{17}{6}sy_t^2 
\right]\,,\\
\beta_{g_2} \eqn{=} 
-\frac{39-s}{12}g_2^3+\frac{1}{16 \pi^2}g_2^3\left[\frac{9}{10}g_1^2
+\frac{35}{6}g_2^2+12g_3^2-\frac{3}{2}sy_t^2\right]\,,\\
\beta_{g_3} \eqn{=} 
-7g_3^3 
+\frac{1}{16 \pi^2}g_3^3\left[\frac{11}{10}g_1^2+\frac{9}{2}g_2^2-26g_3^2-2sy_t^2 
\right]\,.
\end{eqnarray}
The $\beta$-functions for the top Yukawa and the Higgs are given by
\begin{eqnarray}
\beta_{y_t} \eqn{=} 
y_t\left[\left(\frac{23}{6}+\frac{2}{3}s\right)y_t^2
-\left(\frac{17}{20}g_1^2+\frac{9}{4}g_2^2+8g_3^2\right)\right]\notag\\
\eqn{}+\frac{1}{16\pi^2}y_t 
\bigg[ 
-12s^2y_t^4+6s^2\lambda^2-12s^3\lambda y_t^2+\frac{393}{80}sg_1^2y_t^2
+\frac{225}{16}sg_2^2y_t^2+36sg_3^2y_t^2\notag\\
\eqn{}+\frac{1187}{600}g_1^4-\frac{23}{4}g_2^4-108g_3^4-\frac{9}{20}g_1^2g_2^2 
+9g_2^2g_3^2+\frac{19}{9}g_1^2g_3^2 \bigg]\,,
\end{eqnarray}
\begin{eqnarray}
\beta_\lambda \eqn{=} 
6(1+3s^2)\lambda^2-6y_t^4+12\lambda y_t^2
-3\lambda\left(\frac{3}{5}g_1^2+3g_2^2\right)
+\frac{3}{8}\left[2g_2^4+\left(\frac{3}{5}g_1^2+g_2^2\right)^2\right] \notag\\
\eqn{}+\frac{1}{16 \pi^2}\bigg[ 
-(48+288s-324s^2+624s^3-324s^4)\lambda^3 \notag\\
\eqn{}
+\lambda^2\left[\frac{3}{5}(9+18s+9s^2)g_1^2+(27+54s+27s^2)g_2^2\right] \notag\\
\eqn{}
-\lambda\left[-\frac{90+377s+162s^2}{24}\frac{9}{25}g_1^4
-\frac{3-18s+9s^2}{4}\frac{3}{5}g_1^2g_2^2+\frac{181+54s+27s^2}{8}g^4\right]
\notag\\
\eqn{}
+\frac{912+3s}{48}g_2^6-\frac{290-s}{48}\frac{3}{5}g_1^2g_2^4
-\frac{560-s}{48}\frac{9}{25}g_1^4g_2^2-\frac{380-s}{48}\frac{27}{125}g_1^6
-32g_3^2y_t^4-\frac{8}{5}g_1^2y_t^4 \notag\\
\eqn{}
-\frac{9}{4}g_2^4y_t^2+\lambda y_t^2\left(\frac{17}{2}g_1^2+\frac{45}{2}g_2^2
+80g_3^2\right)+\frac{3}{5}g_1^2y_t^2\left(-\frac{19}{4}\frac{3}{5}g_1^2
+\frac{21}{2}g_2^2\right)\notag\\
\eqn{}
-(36+108s^2)\lambda^2y_t^2-(12-117s+108s^2)\lambda y_t^4+(38-8s)y_t^6\bigg] \,,
\end{eqnarray}
where we take the renormalization scale as $\mu=\phi$. We do not take into 
account the running effect of $\xi$ as we fix its value during inflation to 
generate density perturbations of the right magnitude, assuming its running 
effects on other coupling constants in SM are sufficiently small. Thus, the 
value of $\xi$ shown in our analysis should be understood as those evaluated 
during inflation. In addition, the cut-off in our model is field-dependent one 
and the model remains in a weak coupling regime during inflation like the 
ordinary Higgs inflation model (see e.g. Ref.~\cite{Bezrukov:2011sz}). Thus, the
 RGEs can be safely used from the low energy up to the inflationary scale.

$s$ is a factor which should be multiplied to all the 
loop-lines of scalar (Higgs) field $\phi$ ($h$). The factor $s$ is given by
\begin{eqnarray}
s=\left(\frac{d\hat{\phi}}{d\phi}\right)^{-2}=\frac{1}{1+\xi\phi^2}.
\end{eqnarray}
This procedure is similar to the case of Higgs inflation with non-minimal 
coupling~\cite{He:2014ora} but note that the definition of $s$ is different from
 the Higgs inflation with non-minimal coupling. Lastly, we note that a non-zero 
non-minimal coupling is induced by the RGE effects (e.g., 
see~\cite{He:2014ora,Elizalde:1993ew,Elizalde:2014xva,Inagaki:2015eza,Elizalde:2015nya} and
 references therein), but its value is considered to be small so that the 
inflationary prediction of $(n_s,r)$ is hardly modified.



\begin{thebibliography}{99}
\bibitem{Chatrchyan:2013lba}
  G.~Aad {\it et al.}  [ATLAS Collaboration],
  Phys.\ Lett.\ B {\bf 716} (2012) 1
  [arXiv:1207.7214 [hep-ex]].

\bibitem{CMS}
  S.~Chatrchyan {\it et al.}  [CMS Collaboration],
  JHEP {\bf 1306} (2013) 081
  [arXiv:1303.4571 [hep-ex]].

\bibitem{Guth:1980zm}
  A.~H.~Guth,
  Phys.\ Rev.\ D {\bf 23} (1981) 347.

\bibitem{Starobinsky:1980te}
  A.~A.~Starobinsky,
  Phys.\ Lett.\ B {\bf 91} (1980) 99.

\bibitem{Sato:1980yn}
  K.~Sato,
  Mon.\ Not.\ Roy.\ Astron.\ Soc.\  {\bf 195} (1981) 467.

\bibitem{Linde:1981mu}
  A.~D.~Linde,
  Phys.\ Lett.\ B {\bf 108} (1982) 389.

\bibitem{Albrecht:1982wi}
  A.~Albrecht and P.~J.~Steinhardt,
  Phys.\ Rev.\ Lett.\  {\bf 48} (1982) 1220.

\bibitem{CervantesCota:1995tz}
  J.~L.~Cervantes-Cota and H.~Dehnen,
  Nucl.\ Phys.\ B {\bf 442} (1995) 391
  [astro-ph/9505069].

\bibitem{Bezrukov:2007ep}
  F.~L.~Bezrukov and M.~Shaposhnikov,
  Phys.\ Lett.\ B {\bf 659} (2008) 703
  [arXiv:0710.3755 [hep-th]].

\bibitem{Bezrukov:2009db}
  F.~Bezrukov and M.~Shaposhnikov,
  JHEP {\bf 0907} (2009) 089
  [arXiv:0904.1537 [hep-ph]].

\bibitem{Einhorn:2009bh}
  M.~B.~Einhorn and D.~R.~T.~Jones,
  JHEP {\bf 1003} (2010) 026
  [arXiv:0912.2718 [hep-ph]].

\bibitem{Germani:2010gm}
  C.~Germani and A.~Kehagias,
  Phys.\ Rev.\ Lett.\  {\bf 105} (2010) 011302
  [arXiv:1003.2635 [hep-ph]].

\bibitem{Ferrara:2010yw}
  S.~Ferrara, R.~Kallosh, A.~Linde, A.~Marrani and A.~Van Proeyen,
  Phys.\ Rev.\ D {\bf 82} (2010) 045003
  [arXiv:1004.0712 [hep-th]].

\bibitem{Lee:2010hj}
  H.~M.~Lee,
  JCAP {\bf 1008} (2010) 003
  [arXiv:1005.2735 [hep-ph]].

\bibitem{Ferrara:2010in}
  S.~Ferrara, R.~Kallosh, A.~Linde, A.~Marrani and A.~Van Proeyen,
  Phys.\ Rev.\ D {\bf 83} (2011) 025008
  [arXiv:1008.2942 [hep-th]].

\bibitem{Nakayama:2010sk}
  K.~Nakayama and F.~Takahashi,
  JCAP {\bf 1102} (2011) 010
  [arXiv:1008.4457 [hep-ph]].

\bibitem{Kamada:2010qe}
  K.~Kamada, T.~Kobayashi, M.~Yamaguchi and J.~Yokoyama,
  Phys.\ Rev.\ D {\bf 83} (2011) 083515
  [arXiv:1012.4238 [astro-ph.CO]].

\bibitem{Hertzberg:2011rc} 
  M.~P.~Hertzberg,
  JCAP {\bf 1208}, 008 (2012)
  [arXiv:1110.5650 [hep-ph]].

\bibitem{Masina:2011aa}
  I.~Masina and A.~Notari,
  Phys.\ Rev.\ D {\bf 85} (2012) 123506
  [arXiv:1112.2659 [hep-ph]].

\bibitem{Masina:2011un}
  I.~Masina and A.~Notari,
  Phys.\ Rev.\ Lett.\  {\bf 108} (2012) 191302
  [arXiv:1112.5430 [hep-ph]].
  


\bibitem{Kamada:2012se}
  K.~Kamada, T.~Kobayashi, T.~Takahashi, M.~Yamaguchi and J.~Yokoyama,
  Phys.\ Rev.\ D {\bf 86} (2012) 023504
  [arXiv:1203.4059 [hep-ph]].

\bibitem{Allison:2013uaa}
  K.~Allison,
  JHEP {\bf 1402} (2014) 040
  [arXiv:1306.6931 [hep-ph]].

\bibitem{Hamada:2013mya}
  Y.~Hamada, H.~Kawai and K.~y.~Oda,
  PTEP {\bf 2014} (2014) 023B02
  [arXiv:1308.6651 [hep-ph]].

\bibitem{Nakayama:2014koa}
  K.~Nakayama and F.~Takahashi,
  Phys.\ Lett.\ B {\bf 734} (2014) 96;
  [arXiv:1403.4132 [hep-ph]].

\bibitem{Cook:2014dga}
  J.~L.~Cook, L.~M.~Krauss, A.~J.~Long and S.~Sabharwal,
  Phys.\ Rev.\ D {\bf 89} (2014) 10,  103525
  [arXiv:1403.4971 [astro-ph.CO]].

\bibitem{Hamada:2014iga}
  Y.~Hamada, H.~Kawai, K.~y.~Oda and S.~C.~Park,
  Phys.\ Rev.\ Lett.\  {\bf 112} (2014) 24,  241301
  [arXiv:1403.5043 [hep-ph]].

\bibitem{Bezrukov:2014bra}
  F.~Bezrukov and M.~Shaposhnikov,
  Phys.\ Lett.\ B {\bf 734} (2014) 249
  [arXiv:1403.6078 [hep-ph]].

\bibitem{Haba:2014zda}
  N.~Haba and R.~Takahashi,
  Phys.\ Rev.\ D {\bf 89} (2014) 11,  115009
   [Phys.\ Rev.\ D {\bf 90} (2014) 3,  039905]
  [arXiv:1404.4737 [hep-ph]].

\bibitem{Hamada:2014xka}
  Y.~Hamada, H.~Kawai and K.~y.~Oda,
  JHEP {\bf 1407} (2014) 026
  [arXiv:1404.6141 [hep-ph]].

\bibitem{Haba:2014zja}
  N.~Haba, H.~Ishida and R.~Takahashi,
  PTEP {\bf 2015} (2015) 5,  053B01
  [arXiv:1405.5738 [hep-ph]].

\bibitem{He:2014ora}
  H.~J.~He and Z.~Z.~Xianyu,
  JCAP {\bf 1410} (2014) 019
  [arXiv:1405.7331 [hep-ph]].

\bibitem{Hamada:2014raa}
  Y.~Hamada, K.~y.~Oda and F.~Takahashi,
  Phys.\ Rev.\ D {\bf 90} (2014) 9,  097301
  [arXiv:1408.5556 [hep-ph]].

\bibitem{George:2015nza}
  D.~P.~George, S.~Mooij and M.~Postma,
  JCAP {\bf 1604} (2016) no.04,  006
  [arXiv:1508.04660 [hep-th]].

\bibitem{DiVita:2015bha}
  S.~Di Vita and C.~Germani,
  Phys.\ Rev.\ D {\bf 93} (2016) no.4,  045005
  [arXiv:1508.04777 [hep-ph]].
  
\bibitem{Ge:2016xcq}
  S.~F.~Ge, H.~J.~He, J.~Ren and Z.~Z.~Xianyu,
  arXiv:1602.01801 [hep-ph].

\bibitem{Takahashi:2010ky}
  F.~Takahashi,
  Phys.\ Lett.\ B {\bf 693} (2010) 140
  [arXiv:1006.2801 [hep-ph]].

\bibitem{Nakayama:2010kt}
  K.~Nakayama and F.~Takahashi,
  JCAP {\bf 1011} (2010) 009
  [arXiv:1008.2956 [hep-ph]];
  JCAP {\bf 1011} (2010) 039
  [arXiv:1009.3399 [hep-ph]].



\bibitem{Ade:2015lrj}
  P.~A.~R.~Ade {\it et al.} [Planck Collaboration],
  arXiv:1502.02114 [astro-ph.CO].
  
\bibitem{Array:2015xqh} 
  P.~A.~R.~Ade {\it et al.} [BICEP2 and Keck Array Collaborations],
  Phys.\ Rev.\ Lett.\  {\bf 116}, 031302 (2016)
  [arXiv:1510.09217 [astro-ph.CO]].



\bibitem{Bezrukov:2011sz} 
  F.~Bezrukov, D.~Gorbunov and M.~Shaposhnikov,
  JCAP {\bf 1110}, 001 (2011)
  [arXiv:1106.5019 [hep-ph]].

%
\bibitem{ATLAS:2014wva}
  [ATLAS and CDF and CMS and D0 Collaborations],
  arXiv:1403.4427 [hep-ex].

\bibitem{Haba:2014qca}
  N.~Haba, K.~Kaneta, R.~Takahashi and Y.~Yamaguchi,
  Phys.\ Rev.\ D {\bf 91} (2015) 1,  016004
  [arXiv:1408.5548 [hep-ph]].

\bibitem{Horiguchi:2013wra}
  T.~Horiguchi, A.~Ishikawa, T.~Suehara, K.~Fujii, Y.~Sumino, Y.~Kiyo and H.~Yamamoto,
  arXiv:1310.0563 [hep-ex].

\bibitem{Haba:2013lga}
  N.~Haba, K.~Kaneta and R.~Takahashi,
  JHEP {\bf 1404} (2014) 029
  [arXiv:1312.2089 [hep-ph]].

\bibitem{Haba:2014sia}
  N.~Haba, H.~Ishida, K.~Kaneta and R.~Takahashi,
  Phys.\ Rev.\ D {\bf 90} (2014) 036006
  [arXiv:1406.0158 [hep-ph]].
  
\bibitem{Nakayama:2013jka} 
  K.~Nakayama, F.~Takahashi and T.~T.~Yanagida,
  Phys.\ Lett.\ B {\bf 725}, 111 (2013)
  [arXiv:1303.7315 [hep-ph]];
  JCAP {\bf 1308}, 038 (2013)
  [arXiv:1305.5099 [hep-ph]];
  Phys.\ Lett.\ B {\bf 737}, 151 (2014)
  [arXiv:1407.7082 [hep-ph]].

\bibitem{Elizalde:1993ew}
  E.~Elizalde and S.~D.~Odintsov,
  Phys.\ Lett.\ B {\bf 321} (1994) 199
  [hep-th/9311087].

\bibitem{Elizalde:2014xva}
  E.~Elizalde, S.~D.~Odintsov, E.~O.~Pozdeeva and S.~Y.~Vernov,
  Phys.\ Rev.\ D {\bf 90} (2014) no.8,  084001
  [arXiv:1408.1285 [hep-th]].

\bibitem{Inagaki:2015eza}
  T.~Inagaki, S.~D.~Odintsov and H.~Sakamoto,
  Astrophys.\ Space Sci.\  {\bf 360} (2015) no.2,  67
  [arXiv:1509.03738 [hep-th]].

\bibitem{Elizalde:2015nya}
  E.~Elizalde, S.~D.~Odintsov, E.~O.~Pozdeeva and S.~Y.~Vernov,
  JCAP {\bf 1602} (2016) no.02,  025
  [arXiv:1509.08817 [gr-qc]].
  

  
\end{thebibliography}
\end{document}